\documentclass[aps,preprint,nofootinbib,superscriptaddress]{revtex4}
\usepackage{epsfig}

\begin{document}

\date{\today}
\title{Further Development of the Improved QMD Model and its Applications to
Fusion Reaction near Barrier}
\author{Ning Wang}
\email{wangning@itp.ac.cn} \affiliation{China Institute of Atomic
Energy, P. O. Box 275(18), Beijing 102413, P. R. China}
\affiliation{Institute of Theoretical Physics, Chinese Academy of
Sciences, Beijing 100080, P. R. China}
\author{Zhuxia Li}
\email{lizwux@iris.ciae.ac.cn} \affiliation{China Institute of
Atomic Energy, P. O. Box 275(18), Beijing 102413, P. R. China}
\affiliation{Institute of Theoretical Physics, Chinese Academy of
Sciences, Beijing 100080, P. R. China} \affiliation{Nuclear Theory
Center of National Laboratory of Heavy Ion Accelerator, Lanzhou
730000, P. R. China}
\author{Xizhen Wu}
\email{lizwux@iris.ciae.ac.cn}
\affiliation{China Institute of Atomic Energy, P. O. Box 275(18), Beijing 102413, P. R.
China}
\affiliation{Nuclear Theory Center of National Laboratory of Heavy Ion Accelerator,
Lanzhou 730000, P. R. China}
\author{Junlong Tian}
\affiliation{China Institute of Atomic Energy, P. O. Box 275(18), Beijing 102413, P. R.
China}
\author{YingXun Zhang}
\affiliation{China Institute of Atomic Energy, P. O. Box 275(18), Beijing 102413, P. R.
China}
\author{Min Liu}
\affiliation{China Institute of Atomic Energy, P. O. Box 275(18), Beijing 102413, P. R.
China}

\begin{abstract}
The Improved Quantum Molecular Dynamics model is further developed
by introducing new parameters in interaction potential energy
functional based on Skyrme interaction of SkM$^{*}$ and SLy
series. The properties of ground states of selected nuclei can be
reproduced very well. The Coulomb barriers for a series of
reaction systems are studied and compared with the results of the
proximity potential. The fusion excitation functions for a series
of fusion reactions are calculated and the results are in good
agreement with experimental data.
\end{abstract}

\maketitle


\begin{center}
\textbf{I. INTRODUCTION}
\end{center}

Recently, the study of the mechanism for heavy-ion fusion
reactions at energies near the Coulomb barrier, especially, the
mechanism for the enhancement of fusion cross sections for
neutron-rich systems has attracted a lot of attention. The
knowledge of the mechanism of the enhancement of fusion cross
sections for heavy and neutron-rich systems is useful in the
synthesis of superheavy elements. In heavy-ion fusion reactions,
the excitation and deformation of projectile and target, neck
formation and nucleon transfer strongly influence the dynamics of
fusion processes. These effects are more pronounced at near and
below the fusion barrier. For heavy systems, in order to consider
these effects a very large number of degrees of freedom of motion
is involved and the situation becomes very complicated. Thus one
will meet great difficulty by macroscopic dynamics
models\cite{Ada97} in which only few degrees of freedom of motion
are included. The difficulty is also encountered by the fusion
coupled channel model\cite{Hag99} for it is quite difficult even
impossible to include such a large number of possible channels in
practical calculations. Therefore, it is highly requisite to
develop a microscopic dynamical model suitable for studying heavy
ion fusion reactions by which one can consistently take account of
the dynamical deformation, particle transfer, isospin and mass
asymmetry effects, etc.  In our previous work\cite{Wa02,Wa03}, an
improved quantum molecular dynamics (ImQMD) model was proposed.
Main improvements in the ImQMD model are as follows: The surface
and surface symmetry energy terms are introduced in the potential
energy part; A system size dependent wave packet width is
introduced in order to consider the evolution of the wave packet
width; An approximate treatment of anti-symmetrization, namely, a
phase space occupation constraint is adopted \cite{Papa00}. With
this model we have studied the dynamical evolution of the fusion
barrier as well as the development of the neck in fusion reactions
of $^{40,48}$Ca+$^{90,96}$Zr. However, more tests of our model are
needed. Before that one of the most urgent problem has to be
solved is to increase the time of keeping an individual nucleus to
be stable and in a good shape (close to ground state shape)
without emission of nucleons in order to study the dynamical
process for fusion reactions of heavy nuclei. As is well know
 that the time scale for the formation process of a
compound system in  a fusion reaction of heavy nuclei is about
thousands fm/c, and also the time scale of quasi-fission which
reduces the fusion probability substantially in fusion processes
of heavy nuclei is about several thousands fm/c or longer. To meet
this requirement, in this paper we develop an updated version of
improved quantum molecular dynamics model named ImQMD-II based on
our previously work\cite{Wa02,Wa03}. In this paper we mainly
devote to modify the parameters of the potential energy functional
in the model. As is well known that the parameters of the
potential energy functional in the QMD model are obtained based on
Skyrme forces. There are quite a lot of new versions of Skyrme
forces having been developed following the development of the
knowledge of nuclear equation of state. The newly developed Skyrme
forces such as the $SLy$ series are designed to study the
properties of nuclei away from the $\beta $-stability line in
addition to nuclei along the $\beta $-stability line. It has been
shown that these modern Skyrme forces can describe the properties
of the nuclei away from $\beta $-stability line better than the
old Skyrme forces\cite{Chab97}. Therefore it seems to us to be
worthwhile to try new parameters for the ImQMD model based on the
modern parametrizations of Skyrme forces such as the $SLy$ series
as well as the other popular versions.

The paper is organized as follows. In Sec.II, we briefly introduce
the formalism of the ImQMD-II model and the improvements. The
applications of this model to the fusion reactions at energies
near the Coulomb barrier are reported in Sec.III. Finally, the
summary and discussion are given in Sec.IV.

\begin{center}
\textbf{II. FURTHER DEVELOPMENT OF THE IMPROVED QMD MODEL}
\end{center}

In this section we introduce the updated ImQMD model(ImQMD-II) in
more details. First, a brief introduction to the ImQMD model is
presented. Then, we will give the new development of ImQMD.
Finally, the calculation results about the properties of selected
nuclei with the ImQMD-II model are given.

\begin{center}
\textbf{A. Brief Introduction of the ImQMD model}
\end{center}

For readers convenience, let us first briefly introduce the ImQMD model.
In the ImQMD model,the same as in the original QMD model\cite{Hart89,Ai91,Hart98,Ono92},
each nucleon is represented by a coherent state of a Gaussian wave packet
\begin{equation}  \label{1}
\phi _{i}(\mathbf{r})=\frac{1}{(2\pi \sigma _{r}^{2})^{3/4}}\exp [-\frac{(%
\mathbf{r-r}_{i})^{2}}{4\sigma _{r}^{2}}+\frac{i}{\hbar}\mathbf{r}\cdot
\mathbf{p}_{i}],
\end{equation}
where, $\mathbf{r}_{i}, \mathbf{p}_{i}$, are the centers of i-th
wave packet in the coordinate and momentum space, respectively.
$\sigma _{r}$ represents the spatial spread of the wave packet.
The total N-body wave function is assumed to be the direct product
of these coherent states. Through a Wigner transformation, the one-body
 phase space distribution function for N-distinguishable particles is
given by:
\begin{equation}  \label{2}
f(\mathbf{r,p})=\sum\limits_{i}\frac{1}{(\pi\hbar)^{3}}\exp[-\frac{(\mathbf{%
r-r}_{i})^{2}}{2\sigma_{r}^{2}}-\frac{2\sigma_{r}^{2}}{\hbar^{2}}(\mathbf{p-p%
}_{i})^{2}].
\end{equation}

For identical fermions, the effects of the Pauli principle were
discussed in a broader context by Feldmeier and Schnack
\cite{Fe00}. The approximate treatment of anti-symmetrization is
adopted in the ImQMD model by means of the phase space occupation
constraint method \cite{Papa00,Wa02,Wa03}. The density and
momentum distribution functions of a system read
\begin{equation}
\rho (\mathbf{r})=\int f(\mathbf{r,p})d^{3}p=\sum\limits_{i}\rho _{i}(%
\mathbf{r}),  \label{3}
\end{equation}

\begin{equation}
g(\mathbf{p})=\int f(\mathbf{r,p})d^{3}r=\sum\limits_{i}g_{i}(\mathbf{p}),
\label{4}
\end{equation}
respectively, where the sum runs over all particles in the system.
$\rho _{i}(\mathbf{r})$ and $g_{i}(\mathbf{p})$ are the density
and momentum distributions of nucleon i:
\begin{equation}
\rho _{i}(\mathbf{r})=\frac{1}{(2\pi \sigma _{r}^{2})^{3/2}}\exp [-\frac{(
\mathbf{r-r}_{i})^{2}}{2\sigma _{r}^{2}}],  \label{5}
\end{equation}

\begin{equation}  \label{6}
g_{i}(\mathbf{p})=\frac{1}{(2\pi \sigma _{p}^{2})^{3/2}}\exp [-\frac{(%
\mathbf{p-p}_{i})^{2}}{2\sigma _{p}^{2}}],
\end{equation}
where $\sigma_{r}$ and $\sigma_{p}$ are the widths of wave packets
in coordinate and momentum space, respectively, and they satisfy
the minimum uncertainty relation:
\begin{equation}  \label{7}
\sigma_{r}\sigma_{p}=\frac{\hbar}{2}.
\end{equation}

The propagation of nucleons under the self-consistently generated
mean field is governed by Hamiltonian equations of motion:
\begin{equation}  \label{8}
\dot{\mathbf{r}}_{i}=\frac{\partial H}{\partial \mathbf{p}_{i}}, \dot{%
\mathbf{p}}_{i}=-\frac{\partial H}{\partial \mathbf{r}_{i}}.
\end{equation}

The Hamiltonian H consists of the kinetic energy and the effective interaction
potential energy:
\begin{equation}  \label{9}
H=T+U,
\end{equation}
\begin{equation}  \label{10}
T=\sum\limits_{i} \frac{\mathbf{p}_{i}^{2}}{2m}.
\end{equation}

The effective interaction potential energy includes the nuclear local
interaction potential energy and the Coulomb interaction potential energy:
\begin{equation}
U=U_{loc}+U_{Coul}.  \label{11}
\end{equation}%
And%
\begin{equation}
U_{loc}=\int V_{loc}(\mathbf{r})d\mathbf{r},  \label{12}
\end{equation}%
where $V_{loc}(\mathbf{r})$ is potential energy density.

The potential energy density $V_{loc}(\mathbf{r})$ in the ImQMD model reads
\begin{equation}
V_{loc}=\frac{\alpha }{2}\frac{\rho ^{2}}{\rho _{0}}+\frac{\beta }{\gamma +1}%
\frac{\rho ^{\gamma +1}}{\rho _{0}^{\gamma }}+\frac{g_{sur}}{2\rho _{0}}%
(\nabla \rho )^{2}+g_{\tau }\frac{\rho ^{\eta +1}}{\rho _{0}^{\eta }}+\frac{%
C_{s}}{2\rho _{0}}(\rho ^{2}-\kappa _{s}(\nabla \rho )^{2})\delta ^{2},
\label{13}
\end{equation}%
where  $\delta =\frac{\rho _{n}-\rho _{p}}{ \rho _{n}+\rho _{p}}$.
The first three terms in above expression can be obtained from the
potential energy functional of Skyrme forces directly. The fifth
term is the symmetry potential energy part where both the bulk and
the surface symmetry energy are included. In addition, we
introduce an extra small correction term $V_{\tau }=g_{\tau
}\frac{\rho ^{\eta +1}}{\rho _{0}^{\eta }}$ (named tau term) in
the potential energy functional. Inserting expression (13) into
(12), we obtain the local interaction potential energy omitting
self-energies:
\begin{eqnarray}
U_{loc} &=&\frac{\alpha }{2}\sum\limits_{i}\sum\limits_{j\neq i}\frac{\rho
_{ij}}{\rho _{0}}+\frac{\beta }{\gamma +1}\sum\limits_{i}\left(
\sum\limits_{j\neq i}\frac{\rho _{ij}}{\rho _{0}}\right) ^{\gamma }
\label{14} \\
&&+\frac{g_{0}}{2}\sum\limits_{i}\sum\limits_{j\neq i}f_{sij}\frac{\rho _{ij}}{%
\rho _{0}}+g_{\tau }\sum\limits_{i}\left( \sum\limits_{j\neq i}\frac{\rho
_{ij}}{\rho _{0}}\right) ^{\eta }+\frac{C_{s}}{2}\sum\limits_{i}\sum%
\limits_{j\neq i}t_{i}t_{j}\frac{\rho _{ij}}{\rho _{0}}\left( 1-\kappa
_{s}f_{sij}\right),  \nonumber
\end{eqnarray}%
where
\begin{equation}
\rho _{ij}=\frac{1}{(4\pi \sigma _{r}^{2})^{3/2}}\exp [-\frac{(\mathbf{r}_{i}%
\mathbf{-r}_{j})^{2}}{4\sigma _{r}^{2}}],  \label{15}
\end{equation}%
\begin{equation}
f_{sij}=\frac{3}{2\sigma _{r}^{2}}-\left( \frac{\mathbf{r}_{i}\mathbf{-r}_{j}}{%
2\sigma _{r}^{2}}\right) ^{2},  \label{16}
\end{equation}%
and $t_{i}=1$ for protons and $-1$ for neutrons. One should notice
that the third term in (14) comes from both surface term and the
correction to the second term of (13)(see\cite{Wa02}), and thus
$g_{0}$ is actually treated as a parameter in the model.

The Coulomb energy can be written as a sum of the direct and the
exchange contribution, and the latter being taken into account in
the Slater approximation \cite{Slat51,Titi74,Bart02}:
\begin{equation}
U_{Coul}=\frac{1}{2}\int \rho _{^{p}}(\mathbf{r})\frac{e^{2}}{|\mathbf{r-r}%
^{\prime }|}\rho _{^{p}}(\mathbf{r}^{\prime })d\mathbf{r}d\mathbf{r}^{\prime
}-e^{2}\frac{3}{4}\left( \frac{3}{\pi }\right) ^{1/3}\int \rho _{p}^{4/3}d%
\mathbf{R}.  \label{17}
\end{equation}%
where $\rho _{p}$ is the density distribution of protons of the
system. The collision term and phase space occupation constraint
can also readjust the momenta, but the former plays a very small
role in low energy heavy ion collisions and the latter only
happens occasionally. The phase space occupation constraint
method\cite{Papa00} and the system-size-dependent wave-packet
width are adopted as that did in the previous version of
ImQMD\cite{Wa02,Wa03}.

\begin{center}
\textbf{B. The new development in ImQMD(ImQMD-II) model}
\end{center}

The new development of the ImQMD model mainly are reconsidering the parameters
in the potential energy functional (see expression (13)). In the expression
(13), The first three terms can be obtained from the potential energy
functional of a standard Skyrme interaction directly. The parameters $\alpha $, $\beta $%
, $\gamma $ and $g_{sur}$ can be related to the parameters of Skyrme
interactions by
\begin{equation}
\frac{\alpha }{2}\frac{1}{\rho _{0}}=\frac{3}{8}t_{0},  \label{18}
\end{equation}%
\begin{equation}
\frac{\beta }{\gamma +1}\frac{1}{\rho _{0}^{\gamma
}}=\frac{1}{16}t_{3}, \label{19}
\end{equation}
and
\begin{equation}
\frac{g_{sur}}{2\rho _{0}}=\frac{1}{64}(9t_{1}-5t_{2}-4t_{2}x_{2}).
\label{20}
\end{equation}%
Where the parameter $\gamma $ is taken the same value as in the
Skyrme interaction. The linear density dependence of the symmetry
energy term is taken and the parameter is fixed by the symmetry
energy coefficient. In Table.1 we list
the $\alpha $, $\beta $, and $\gamma $ parameters used in the QMD model\cite%
{Ai91} and the corresponding values obtained from various versions
of Skyrme interaction\cite{Bei75,Kri80,Bert82,Dob84,Chab97}. From
the table one can find the parameters of QMD(hard) is close to
that of SIII, and the soft one is close to SkP.
\begin{center}
\text{Table 1}
\end{center}
\begin{center}
\begin{tabular}{lllcl}
\hline\hline
Parameter & $\alpha (MeV)$ & $\beta (MeV)$ & $\ \ \gamma $ \ \  & $%
g_{sur}(MeVfm^{2})$ \\ \hline
QMD(hard) & \multicolumn{1}{c}{-124.0} & \multicolumn{1}{c}{71.0} & 2 &
\multicolumn{1}{c}{-} \\
QMD(soft) & \multicolumn{1}{c}{-356.0} & \multicolumn{1}{c}{303.0} & 7/6 &
\multicolumn{1}{c}{-} \\ \hline
SIII\cite{Bei75} & \multicolumn{1}{c}{-139.6} & \multicolumn{1}{c}{71.4} & 2
& \multicolumn{1}{c}{20.14} \\
SkP\cite{Dob84} & \multicolumn{1}{c}{-362.8} & \multicolumn{1}{c}{309.6} &
7/6 & \multicolumn{1}{c}{19.84} \\
SkM\cite{Kri80} & \multicolumn{1}{c}{-327.3} & \multicolumn{1}{c}{258.0} &
7/6 & \multicolumn{1}{c}{20.32} \\
SkM*\cite{Bert82} & \multicolumn{1}{c}{-327.3} & \multicolumn{1}{c}{258.0} &
7/6 & \multicolumn{1}{c}{21.82} \\
SLy10\cite{Chab97} & \multicolumn{1}{c}{-310.2} & \multicolumn{1}{c}{228.8}
& 7/6 & \multicolumn{1}{c}{21.57} \\ \hline\hline
\end{tabular}
\end{center}

The SIII parametrization was proposed in 1975 and with it one
could describe the ground state properties of spherical nuclei
very well. However, the
incompressibility modulus of symmetric nuclear matter obtained from SIII ($%
K_{\infty }\approx 365MeV$) is too high\cite{Blai80}. Taking this
into account, Krivine et al. derived the SkM interaction at 1980.
Later on, detailed studies of the fission barriers\cite{Bjo80} in
the actinide region resulted in a more refined value of the
nuclear surface tension. Then SkM$^{*}$ \cite{Bert82} was derived
which gave a more refined surface tension. Nowadays, the symmetry
energy part of the interaction attracts a lot interests as nuclei
further away from the stability line can be expected to be
produced with the coming radioactive beam facilities. Putting more
emphases on the isospin degree of freedom, a series of sets of SLy
parametrizations were proposed at the late of 1990's for
reproducing the properties of the nuclei from the $\beta$
stability line to the drip lines \cite{Chab97}.

Considering the successes of SkM$^{\ast }$ in describing the
surface tension and SLy in describing the properties of nuclear
systems far away from $\beta $ stability line, the SkM$^{\ast }$
and SLy parametrization can provide us with a reference to adjust
the new ImQMD parameters. Concerning the symmetry energy term, the
linear density dependence of the bulk symmetry energy term is
taken and the parameter is fixed by the symmetry energy
coefficient. The surface symmetry energy term is also introduced.
This term is important for having a correct neutron skin, which
was introduced in the liquid-drop model\cite{Moll95,Dan03}. For
Skyrme interactions \cite{Vau72,Chab97}, the surface-symmetry term
can be extracted and reads
\begin{equation}
U_{surf-symm}=-\frac{C_{s}\kappa _{s}}{2\rho _{0}}\int [\nabla \rho (\mathbf{
r})]^{2}\delta(\mathbf{r}) ^{2}d\mathbf{r.}  \label{21}
\end{equation}
It modifies the symmetry potential at surface region, and
therefore it is especially important for correctly describing the
neck dynamics in fusion reactions of neutron-rich nuclei. In our
previous work this term was introduced and it was found that this
term played a role in the fusion dynamics. We will study this
effect in the following section. Taking SkM$^{\ast }$ and SLy
parametrizations as the reference we propose a new set of
parameters for the ImQMD model by
reproducing the properties of ground state of selected nuclei $^{208}$Pb,$%
^{90}$Zr,$^{40}$Ca and $^{16}$O and the fusion cross sections of $^{40}$Ca+$%
^{48}$Ca\cite{Tro01}, $^{40}$Ca+$^{90,96}$Zr\cite{Timm98}. The new set of
ImQMD parameters named IQ1 is listed in Table.2

\begin{center}
\text{Table 2}
\end{center}

\begin{tabular}{cccccccccc}
\hline\hline
Para. & $\alpha (MeV)$ & $\beta (MeV)$ & $\gamma $ & $g_{0}(MeVfm^{2})$ & $%
g_{\tau }(MeV)$ & $\eta $ & $C_{s}(MeV)$ & $\kappa _{s}(fm^{2})$ & $\rho
_{0}(fm^{-3})$ \\ \hline
IQ1 & -310.0 & 258.0 & 7/6 & 19.8 & 9.5 & 2/3 & 32.0 & 0.08 & 0.165 \\
\hline\hline
\end{tabular}

\bigskip

From Table 1 and Table 2, one can see that the parameter set of IQ1 is generally
close to SLy10 and SkM*.

\begin{center}
\text{Table 3}
\end{center}
\bigskip
\begin{tabular}{l|l|ccccccccc}
\hline\hline
\multicolumn{2}{c|}{Nuclei \ \ } & $\ ^{208}$Pb \ \  & $\ ^{140}$Ce \ \  & $%
\ ^{132}$Sn\ \  & $\ ^{114}$Sn \ \  & $\ ^{90}$Zr \ \  & $\ ^{56}$Ni \ \  & $%
\ ^{48}$Ca \ \  & $\ ^{40}$Ca \ \  & $\ ^{16}$O \ \  \\ \hline
Binding & exp.\cite{Bert82}$(MeV)$ & 7.87 & 8.38 & 8.35 & 8.53 & 8.71 & 8.64
& 8.67 & 8.55 & 7.98 \\ \cline{2-11}
Energy & IQ1$(MeV)$ & 7.77 & 8.35 & 8.27 & 8.51 & 8.71 & 8.63 & 8.67 & 8.65
& 8.23 \\ \hline
RMS & exp.\cite{Bert82}$(fm)$ & 5.50 & 4.88 &  &  & 4.27 & 3.75 & 3.48 & 3.49
& 2.73 \\ \cline{2-11}
Radius & IQ1$(fm)$ & 5.51 & 4.87 & 4.79 & 4.55 & 4.25 & 3.71 & 3.54 & 3.44 &
2.72 \\ \hline\hline
\end{tabular}

\bigskip
\[
\fbox{Fig. 1}
\]

In Table 3 we list the binding energies and the root-mean-square charge radii of $%
^{208}$Pb, $^{140}$Ce, $^{132}$Sn, $^{114}$Sn, $^{90}$Zr, $^{56}$Ni, $^{48}$%
Ca, $^{40}$Ca, and $^{16}$O calculated by the ImQMD-II model with
parameter set IQ1. The experimental data are also listed for
comparison and it is shown that the calculated results are in good
agreement with experimental data. In Fig.1, we present the time
evolution of binding energies and root-mean-square charge radii
for $^{90}$Zr and $^{208}$Pb calculated by the ImQMD-II model with
IQ1 parameters. One can see that their binding energies, and root
mean square charge radii remain constants with a very small
fluctuation and the bound nuclei evolve stably without spurious
emission for a period of time of about 3000fm/c, which is
essential for applications to fusion reactions of heavy nuclei as
is discussed in the introduction.

\begin{center}
\textbf{III. Applications to fusion Reactions near the Coulomb Barrier}
\end{center}

In this section, we show the calculation results of Coulomb barriers and
fusion excitation functions for a series of fusion systems by means of the
ImQMD-II model with parameters of IQ1 .

\begin{center}
\textbf{A. The Coulomb Barrier}
\end{center}

The interaction potential $V(R)$ is defined by
\begin{equation}
V(R)=E_{12}(R)-E_{1}-E_{2}.  \label{22}
\end{equation}%
Here $R$ is the distance between the centers of mass of projectile
and target. $E_{12}(R)$ is the total energy of whole system, while
$E_{1}$ and $E_{2}$ are the energies of projectile(like) and
target (like) part, respectively. For kinetic energies, the
Thomas-Fermi approximation is adopted as mentioned in
ref.\cite{Denis02}. By using the ImQMD model, both the static and
dynamic Coulomb barrier can be calculated. For the static Coulomb
barrier case, the static density distribution which is the same as
the initial density distribution of projectile and target is
adopted, while for the dynamic Coulomb barrier case the density
distribution of the system changes dynamically due to the
interaction between the reaction partners.

We show the static Coulomb barriers calculated by the ImQMD-II model with parameter set of IQ1
 for $^{40}$Ca+$^{48}$Ca, $^{40}$Ca+$^{90}$%
Zr, $^{16}$O+$^{16}$O and $^{16}$O+$^{208}$Pb in Fig.2. The
results calculated from proximity potential \cite{Myers00} are
also shown in the figures. One can see that the Coulomb barriers
calculated with ImQMD-II are in good agreement with those from
proximity potential\cite{Myers00}.

\[
\fbox{Fig. 2}
\]

\[
\fbox{Fig. 3}
\]%
In addition, the effects of the mass asymmetry of projectile and
target on the static Coulomb barrier are studied through
calculating the static Coulomb barriers for $^{131}$I+$^{131}$I,
$^{54}$Cr+$^{208}$Pb, $^{32}$S+$^{230}$Th and $^{12}$C+$^{250}$Fm
fusion systems which can form the same compound nucleus
$^{262}$Sg. In Fig.3, the solid curves denote the static Coulomb
barriers calculated by ImQMD-II with parameters IQ1 and the
dashed curves denote the results from the proximity potential. From
Fig.3 one can get two points: the first one is that the results
from ImQMD-II are in good agreement with those from proximity
potential when two nuclei do not overlap too much in space. The
proximity potential may not be able to give an accurate result at
the overlapping region where the ImQMD model is applicable.
Clearly, the results for the overlapping region are more
interesting, especially for the cases of heavy systems. The second
one is that with the increase of the mass asymmetry of projectile
and target (from $^{131}$I+$^{131}$I to $^{12}$C+$^{250}$Fm) the
height of the Coulomb barrier decreases gradually and the capture
probability should be enhanced consequently. \cite{Denis02}.

\begin{center}
\textbf{B. Fusion cross sections}
\end{center}

In the ImQMD model, we first create certain reaction events for
each incident energy $E$ and impact parameter $b$ (in tis work the
number is 100) and then counting the number of fusion events, we
obtain the probability of fusion reactions, $g_{fus}(E,b)$, then
the cross section is calculated by using the
expression\cite{Wa02}:
\begin{equation}
\sigma _{fus}(E)=2\pi \int\limits_{0}^{b_{\max }}bg_{fus}(E,b)db=2\pi \sum
bg_{fus}(E,b)\Delta b.  \label{23}
\end{equation}%
At a certain incident energy, the probability of fusion reactions,
$g_{fus}(E,b)$, decreases with the increase of impact parameter as
shown in  Fig.4. This is because the interaction between two
nuclei decreases gradually from central collisions to peripheral
collisions, and following it the probability for fusion reactions
decreases and that for elastic scattering processes increases. In
addition, the $g_{fus}(E,b)$ decreases quickly with the decrease
of incident energies within the energy range interested in this
work. Fig.5 presents the evolution of the fusion probability with
impact parameter at $E_{c.m.}=60,58,54,52MeV$, respectively. One
can see from the figure that the fusion probability falls when the
incident energy decreases from 60MeV to 52MeV and for energies
below the Coulomb barrier, fusion events generally only occur at
central collisions.
\[
\fbox{Fig.4}
\]%
\[
\fbox{Fig.5}
\]%
\[
\fbox{Fig.6}
\]

In Fig.6, we show the fusion excitation functions for $^{40}$Ca+$^{48}$Ti %
\cite{Bier95}, $^{46}$Ti+$^{46}$Ti\cite{Ste00}, $^{40}$Ca+$^{90,96}$Zr\cite%
{Timm98}, $^{32,34}$S+$^{89}$Y\cite{Huk02}, and $^{28}$Si,$^{35}$Cl+$^{92}$%
Zr \cite{New01} at energies near the Coulomb barrier, and the experimental
data are also presented for comparison. In the figure the triangles denote
the results of ImQMD-II with parameters of IQ1 and the circles
denote the experimental data and crosses denote the results of
one-dimension WKB approximation\cite{Wong73}. One can see from the figure
that calculation results of ImQMD-II for the fusion excitation functions are
generally in good agreement with experimental data, which implies that our
model is quite reasonable. Now let
us study the influence of the surface-symmetry energy term on the fusion cross
sections, we show the calculation results of the fusion excitation
functions of $^{40}$Ca+$^{48}$Ca \cite {Tro01} at energies near
the Coulomb barrier without and with the surface-symmetry energy
term taken into account in Fig.7. One can find from the figure that at incident energies above
the Coulomb barrier the fusion cross sections calculated under two cases are
approximately equal, while at incident energies below the barrier the
difference between two cases become obvious, and for this case the fusion cross sections
without surface-symmetry energy term taken into account are obviously larger
than the experimental data and
those with surface-symmetry energy term taken into account
can reproduce the experimental data well. We have
found that the N/Z ratio at neck region is enhanced at the early stage for
neutron-rich fusion process\cite{Wa03}, which is driven by the symmetry
potential. It lowers the fusion barrier and consequently enhances the fusion
cross section. If only the bulk term is taken in account, the effect of the
symmetry energy term becomes too strong at surface region and therefore a
surface-symmetry energy term should be taken into account in order to reduce the
effect of the bulk symmetry energy term at surface region. It is especially
important for having a correct neck dynamics in neutron-rich nuclear fusion
reactions.

\begin{center}
\[
\fbox{Fig.7}
\]%
\[
\fbox{Fig.8}
\]
\end{center}

For further testing the reliability of ImQMD, we calculate the
excitation function for fusion reactions of the neutron-rich
radioactive beam $^{132}$Sn bombarding on the neutron-rich target
$^{64}$Ni at energies near the Coulomb barrier, which was recently
measured \cite{Liang03}. Fig.8 shows the comparison of our
calculation results and the experimental data as well as the
results from coupled-channel calculations \cite{Liang03}. The
solid circles denotes the experimental data, the triangles denote
the results of ImQMD-II with parameters of IQ1. The dashed and
solid curves denote the results of coupled-channel calculations
with including inelastic excitation (IE) of the projectile and
target and IE plus neutron transfer (n\&IE), respectively (see
ref.\cite{Liang03} and references therein). From Fig.8 one can
find that the coupled-channel calculations significantly
under-predicted the sub-barrier fusion cross
sections\cite{Liang03}. This is because with the coupled-channel
model it is difficult to consider all degrees of freedom of
motion, the number of which is extremely large in sub-barrier
fusion reactions for heavier systems, while with the ImQMD model,
all degrees of freedom of motion are self-consistently included.
Thus, this model may possibly provide us with a useful approach to
explore the mechanism of the capture process in the synthesis of
superheavy elements. The work concerning this aspect is in
progress and the results will be reported in the future work.
\[
\fbox{Fig.9}
\]

As a test we also make applications of our model to describe the charge
distributions of fragments in multifragmentation processes. In Fig.9 we show
the charge distribution of fragments by using the ImQMD-II model with parameter set of IQ1
for Ca+Ca at $E=35MeV/nucleon$\cite{Hag94} and Xe+Sn at $E=50MeV/nucleon$%
\cite{Hud03}. One can find that the charge distribution of
fragments, especially the number of intermediate mass fragments is
in good agreement with experimental data. It is well known that
the number of intermediate mass fragments is usually being
under-predicted in QMD model calculations. Our results are
encouraging and it seems to us that the model can be used for
heavy ion collisions at both low energies and intermediate energies.

\begin{center}
\textbf{IV. Conclusions and Discussion}
\end{center}

In this work, we have made further improvements in the ImQMD model
and proposed a new version, namely, the ImQMD-II model in which a
new small correction term and a surface symmetry energy term are
introduced in the potential energy functional in addition to the
terms adopted in the normal isospin dependent QMD model.  A
parameter set for the potential energy functional based on Skyrme
interaction of SkM $^{*}$ and SLy series is introduced. By using
the new version of the ImQMD model, the ground state properties of
a series selected nuclei can be described very well. The time
evolution of individual nuclei can keep stable for about several
thousands fm/c which roughly fits the requirement for study of
fusion reactions of heavy nuclei. We have shown that with this
model both the Coulomb barriers and the fusion excitation
functions for a series of fusion systems (including neutron-rich
radioactive beam $^{132}$Sn +$^{64}$Ni fusion reactions) at
energies near the barrier can be reproduced very well. Our study
shows that the microscopic dynamical model such as the ImQMD model
has an advantage of taking account of the dynamical effects such
as the excitations of projectile and target (deformation and
vibration), neck formation, isospin and mass asymmetry, etc.
simultaneously and thus offers a useful way to study fusion
reactions of heavy nuclei. Furthermore, we have also shown that
this model seems to work well on the charge distribution of
fragments in multifragmentation processes.

\section{Acknowledgements}

This work is supported by National Natural Science Foundation of
China under Grant Nos. 10235030, 10235020, 10175093, 10175089 
and Major State Basic
Research Development Program under Contract, No. G20000774.

\textsf{\newpage }

\begin{center}
\textsf{{\small \textbf{CAPTIONS} } }
\end{center}

\begin{description}
\item[\texttt{Fig.1}] The time evolution of binding energies and
root-mean-square charge radii for $^{90}$Zr and $^{208}$Pb
calculated by the ImQMD-II model with parameter set of IQ1.

\item[\texttt{Fig.2}] The static Coulomb barriers of $^{40}$Ca+$^{48}$Ca, 
$^{40}$Ca+$^{90}$Zr, $^{16}$O+$^{16}$O and $^{16}$O+$^{208}$Pb. 
The solid and dashed curves denote the results of ImQMD-II with parameters of IQ1
and those of proximity potential, respectively.

\item[\texttt{Fig.3}] The static Coulomb barriers of $^{131}$I+$^{131}$I,
 $^{54}$Cr+$^{208}$Pb, $^{32}$S+$^{230}$Th and $^{12}$C+$^{250}$Fm. The
solid and dashed curves denote the results of ImQMD-II with parameters of IQ1
and those of proximity potential, respectively.

\item[\texttt{Fig.4}] The probability of fusion reaction $g_{fus}(E,b)$
as a function of impact parameter for  $^{40}$Ca+$%
^{48}$Ca at $E_{c.m.}=60MeV$.

\item[\texttt{Fig.5}] The probability of fusion reaction $g_{fus}(E,b)$ for $^{40}$Ca+$%
^{48}$Ca at $E_{c.m.}=60,58,54,52MeV$.

\item[\texttt{Fig.6}] The fusion excitation functions of a)$^{40}$Ca+$^{48}$Ti
\cite{Bier95}, b)$^{46}$Ti+$^{46}$Ti\cite{Ste00}, c) and d) $^{40}$Ca+$^{90,96}$Zr\cite%
{Timm98}, e) and f) $^{32,34}$S+$^{89}$Y\cite{Huk02}, and g) and h)$^{28}$Si,$^{35}$Cl+$^{92}$Zr%
\cite{New01} at energies near the Coulomb barrier. The open circles denote
the experimental data, and the filled triangles denote the results of ImQMD-II with
parameters of IQ1. The crosses denote the results of one-dimension WKB approximation

\item[\texttt{Fig.7}] The fusion excitation function for $^{40}$Ca+$^{48}$%
Ca. The stars denote the experimental data. The solid circles and
triangles denote the results without and with the surface-symmetry
term taken into account, respectively.

\item[\texttt{Fig.8}] The fusion excitation function of $^{132}$Sn+$^{64}$Ni%
\cite{Liang03}. The solid circles denote the experimental data.
The filled triangles denote the results of ImQMD-II with parameters of
IQ1. The dashed and solid curves denote the results of
coupled-channel calculations including inelastic excitation (IE)
of the projectile and target and IE plus neutron transfer (n\&IE),
respectively.

\item[\texttt{Fig.9}] The charge distribution of fragments for (a) Ca+Ca at $%
E=35MeV/nucleon$ and (b) Xe+Sn $E=50MeV/nucleon$. The open circles
denote the experimental data, the stars and solid circles denote
the results of ImQMD-II\ model without and with the $g_{\tau}$
term taken into account, respectively. The calculation results are
averaged over the impact parameter range of $b=1-3fm$.

\item[\texttt{Table.1}] parameters used in the QMD model  and the
corresponding values obtained from various Skyrme
parametrizations.

\item[\texttt{Table.2}] The parameters of IQ1.

\item[\texttt{Table.3}] The binding energies and root-mean-square
charge radii of a series ground state nuclei calculated by the
ImQMD-II model with IQ1 interaction.
\end{description}

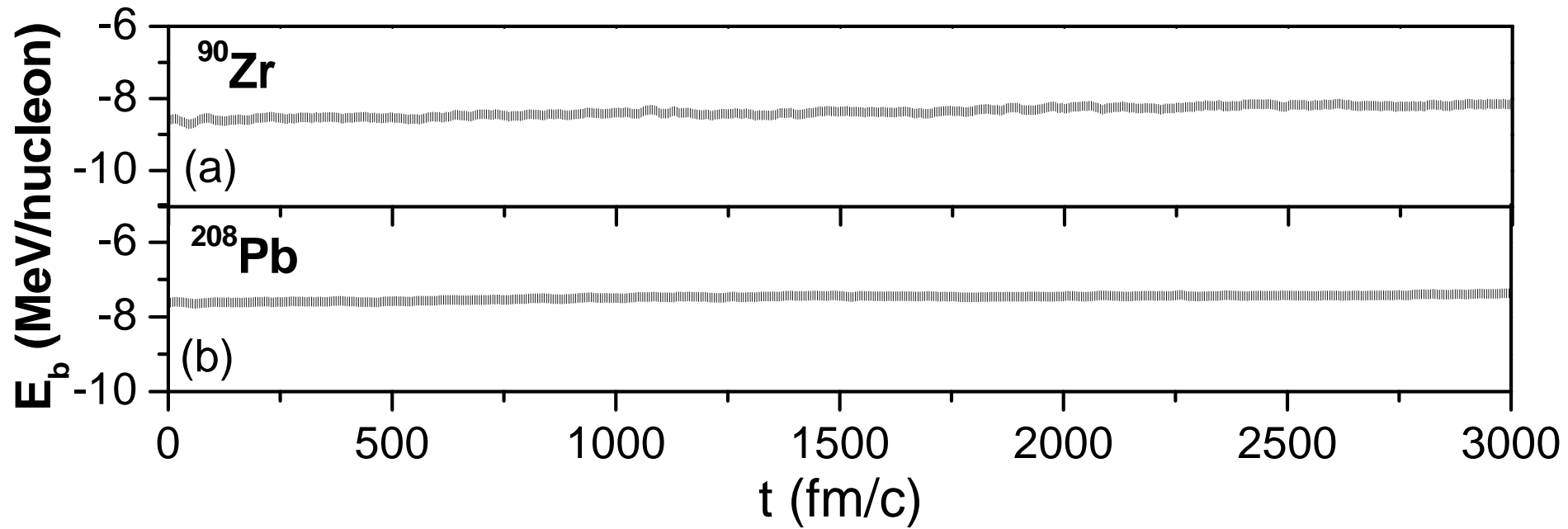
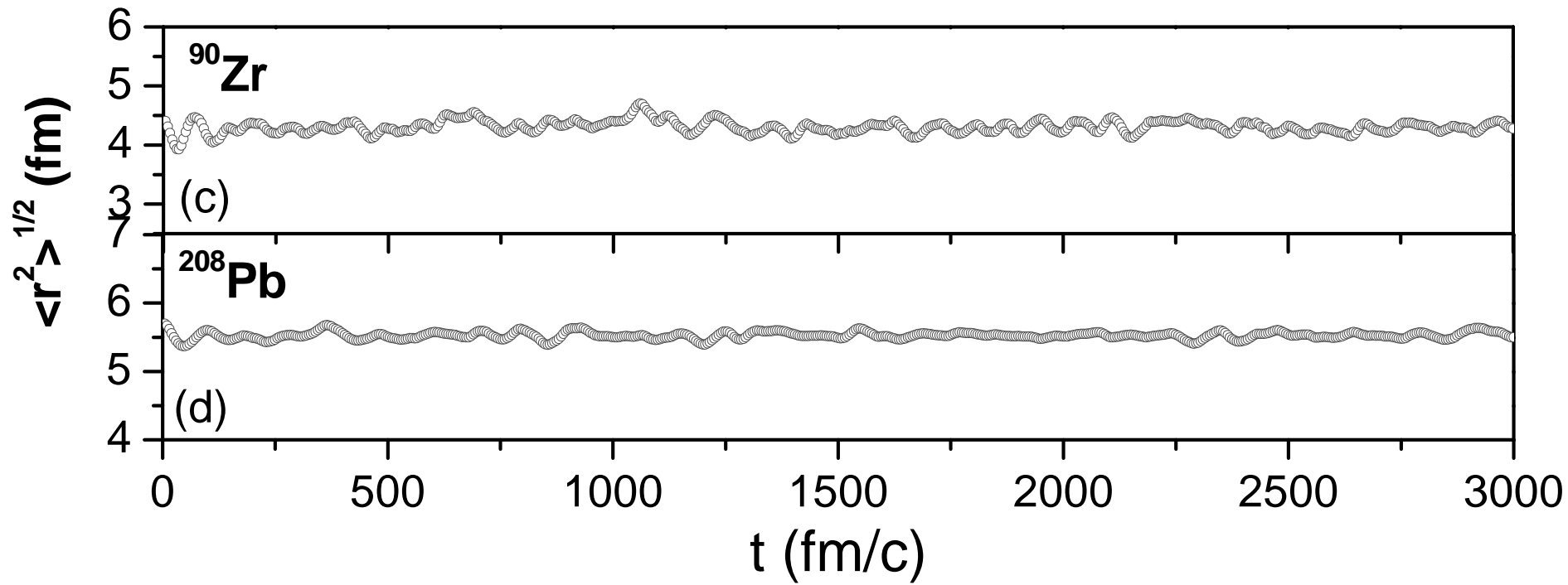
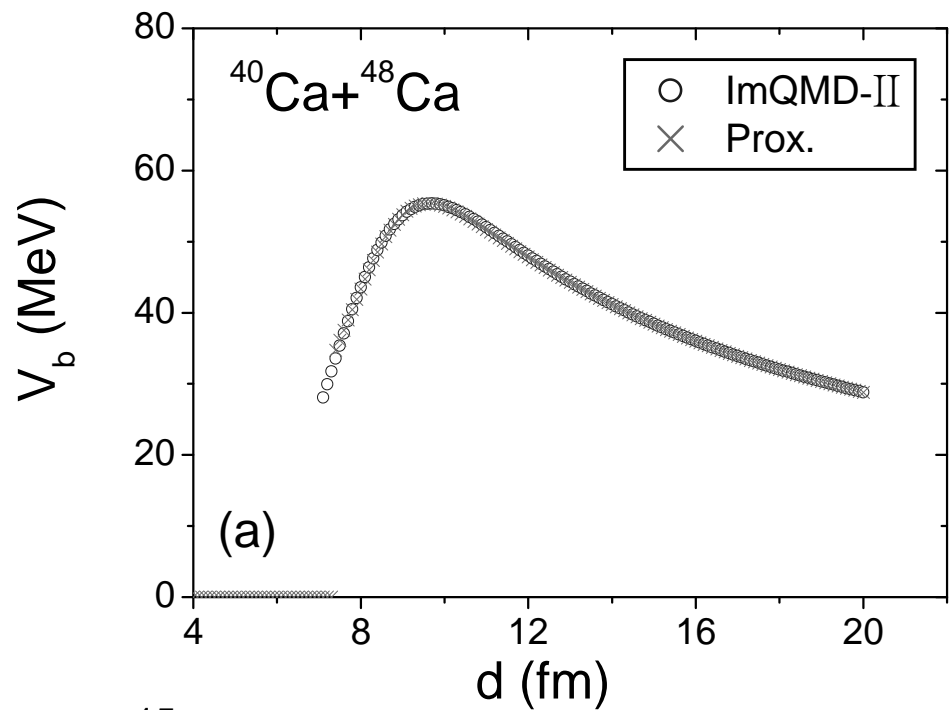
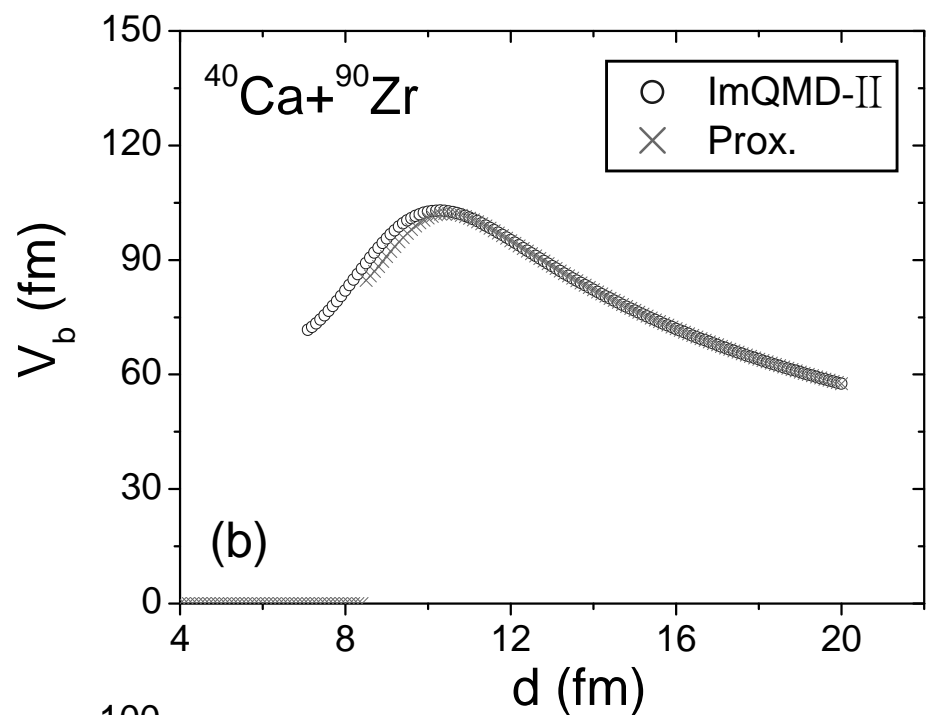
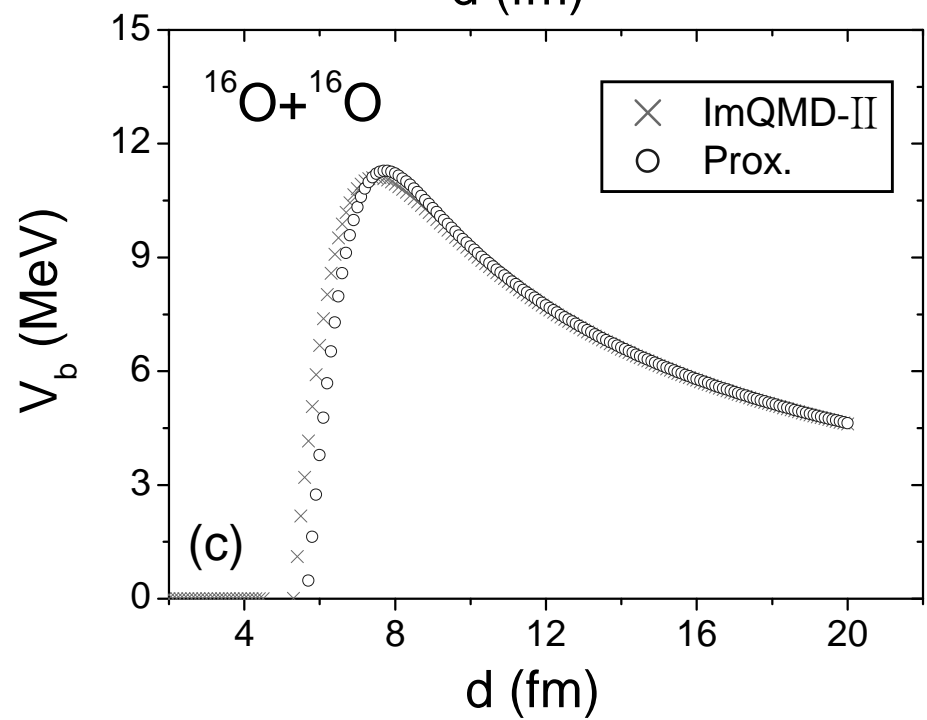
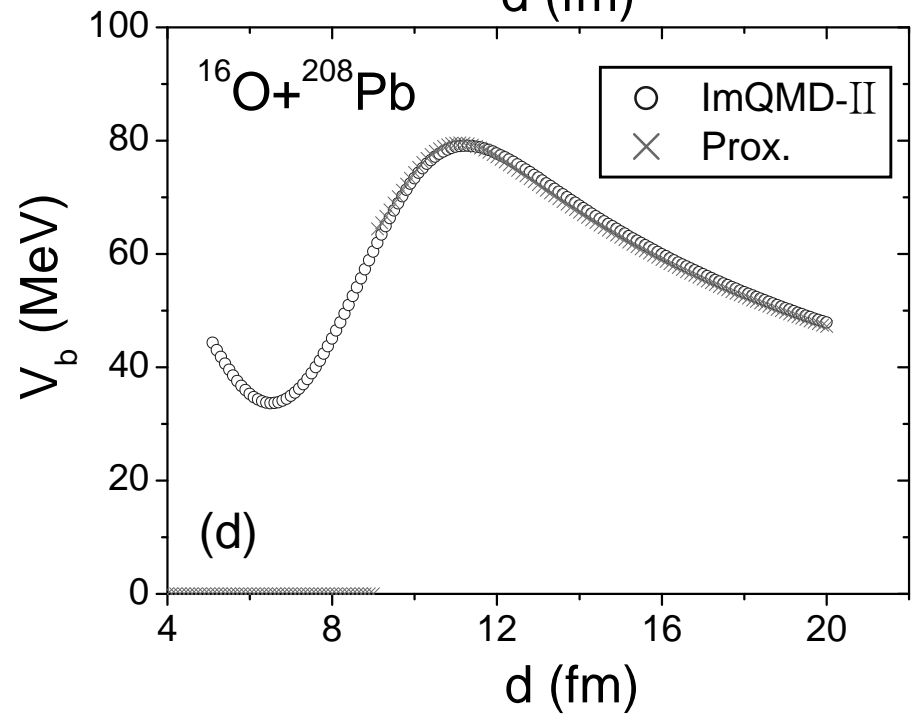
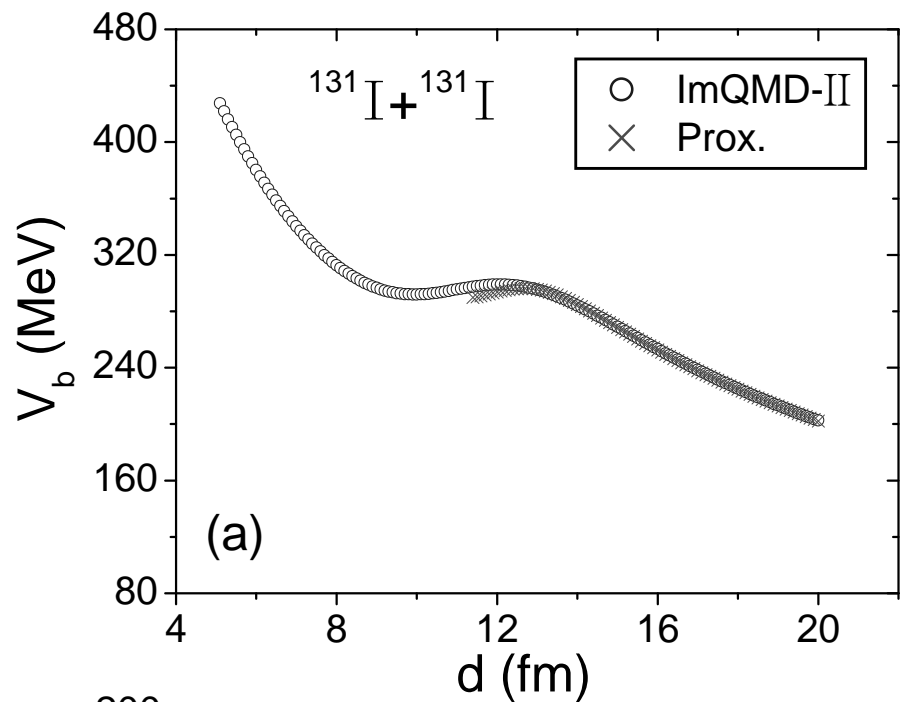
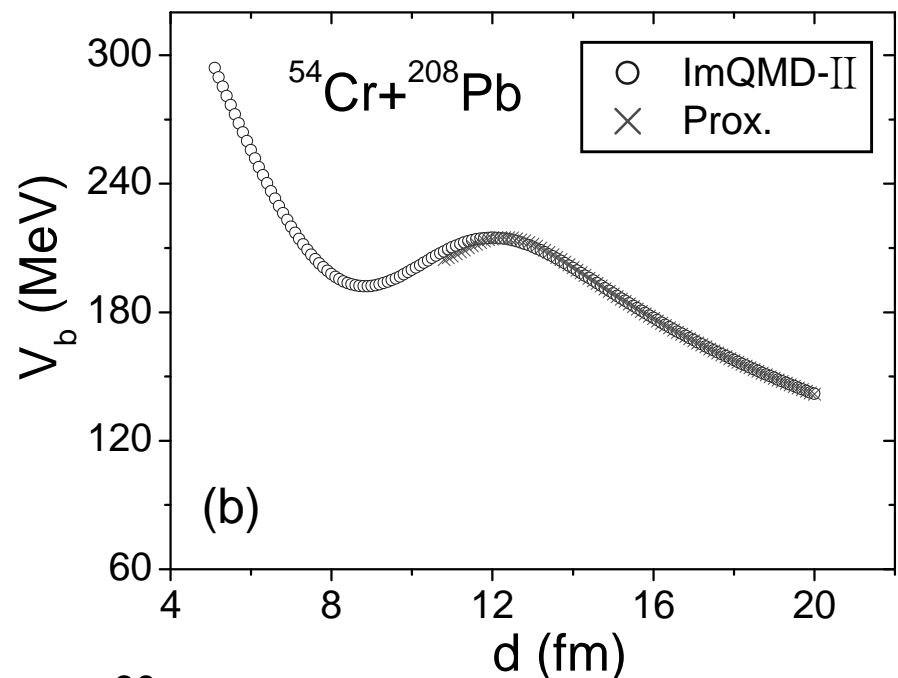
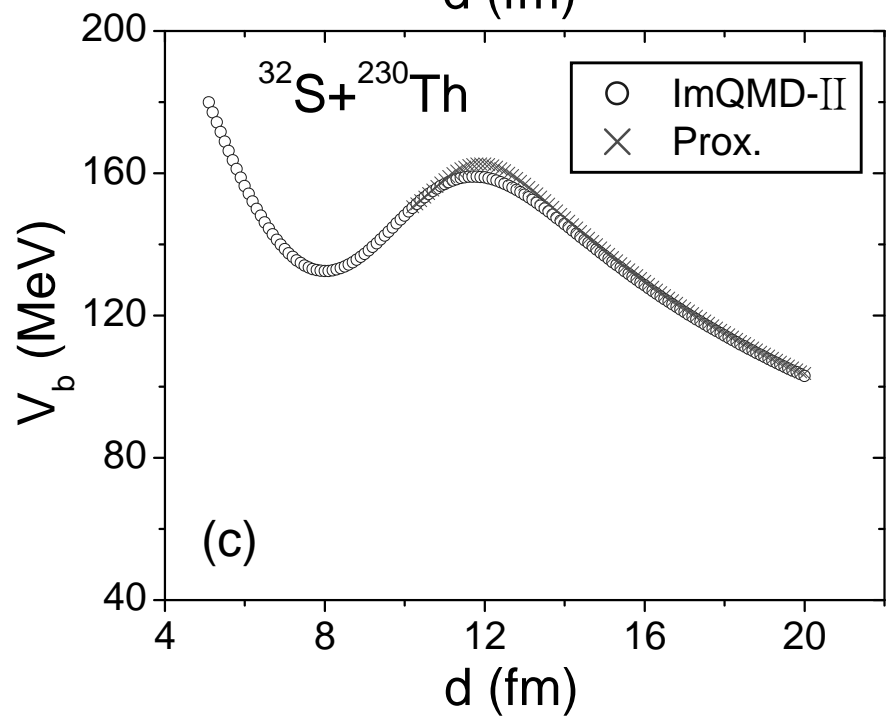
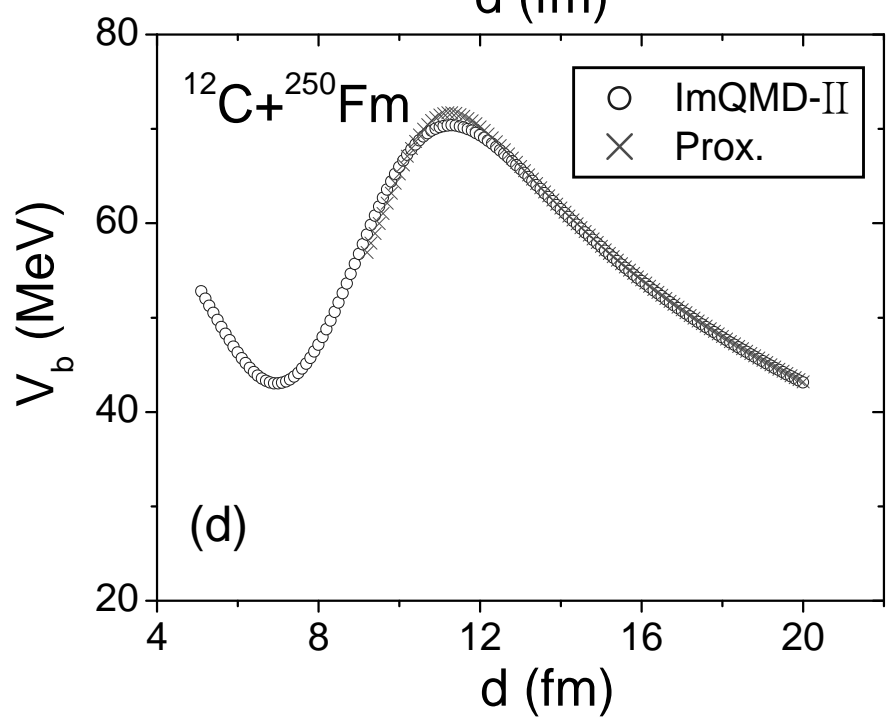
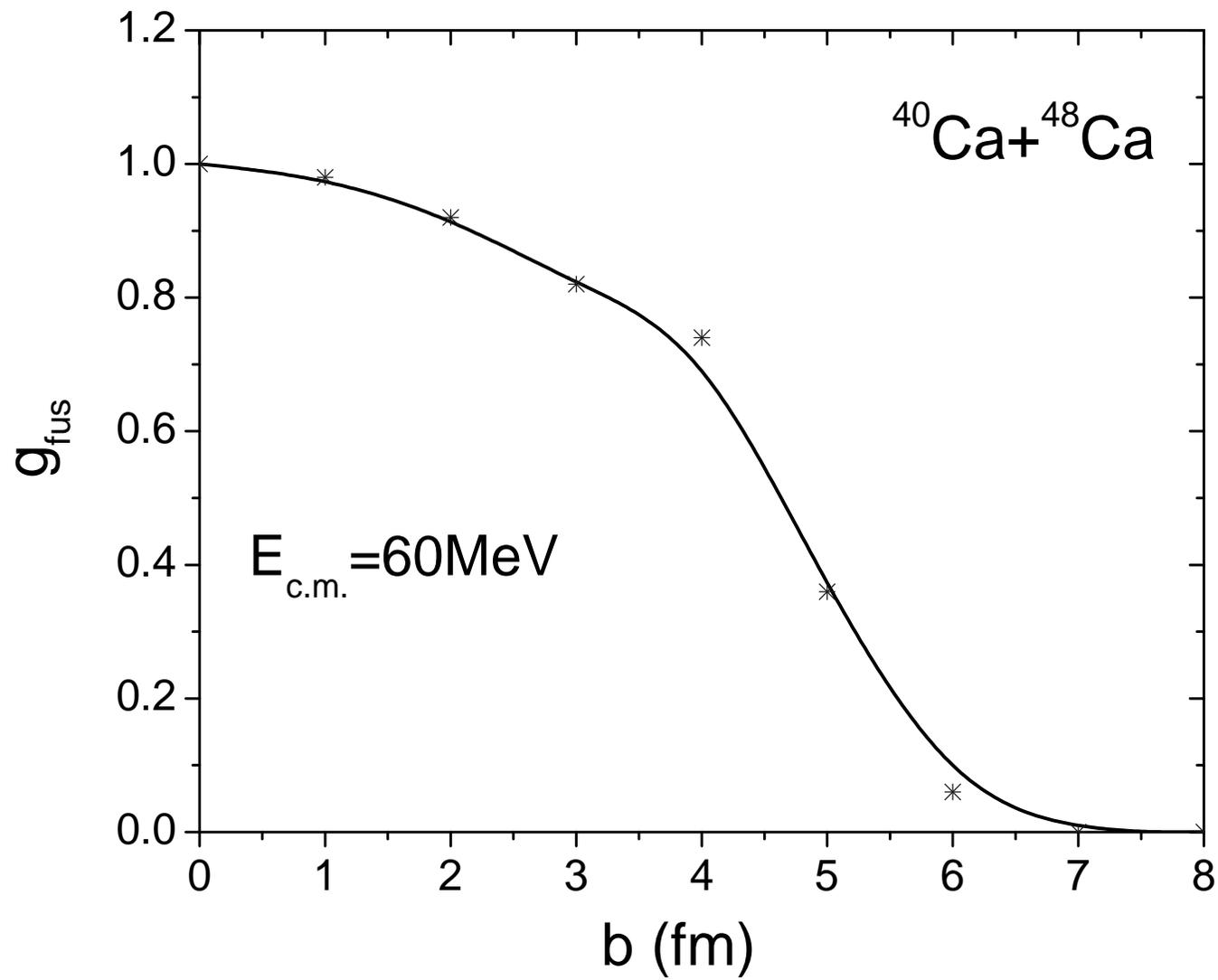
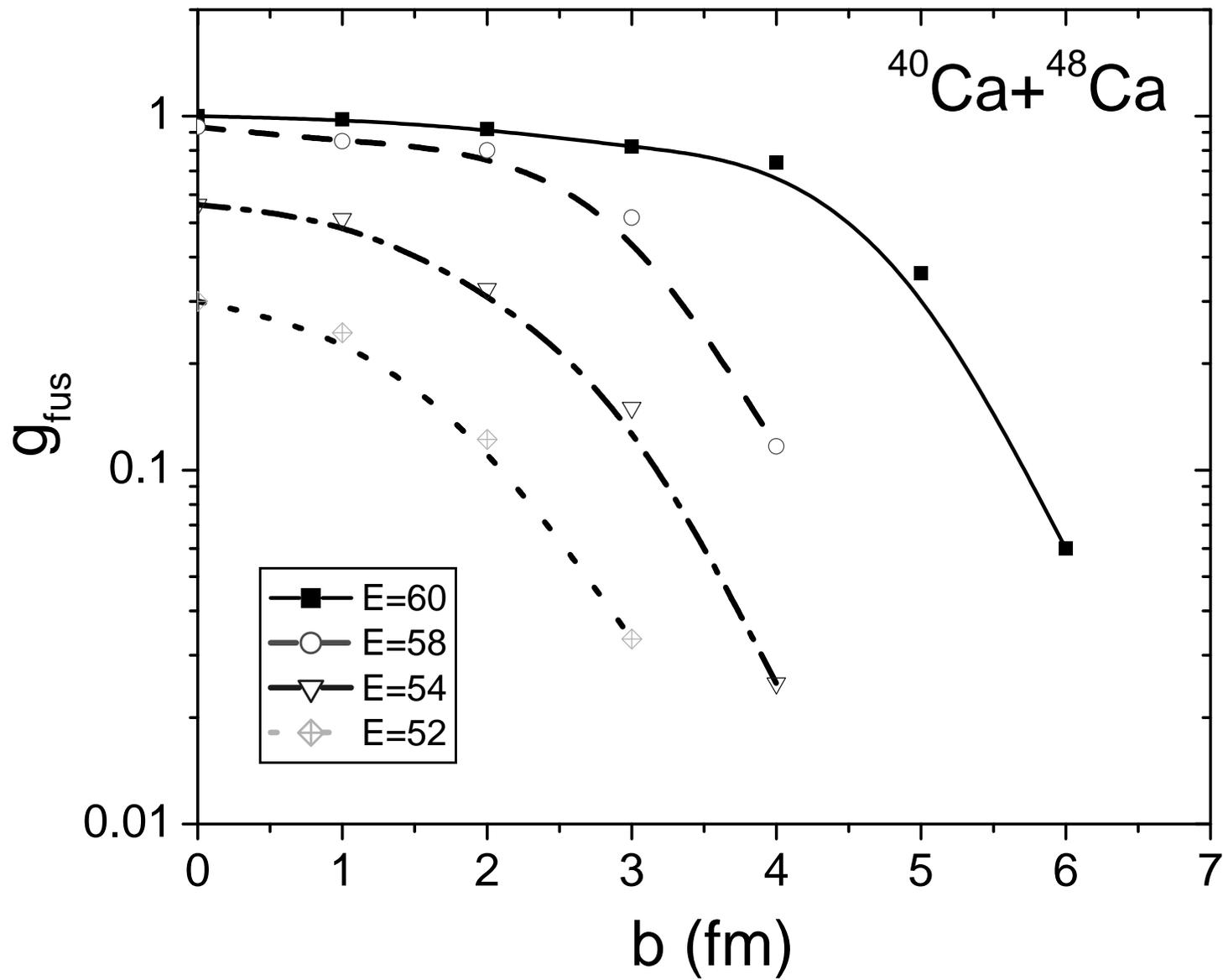
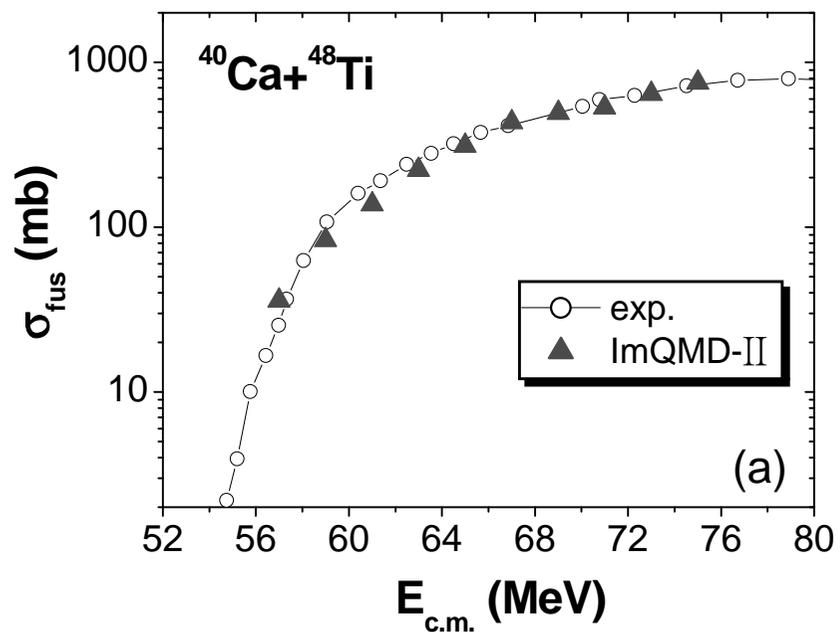
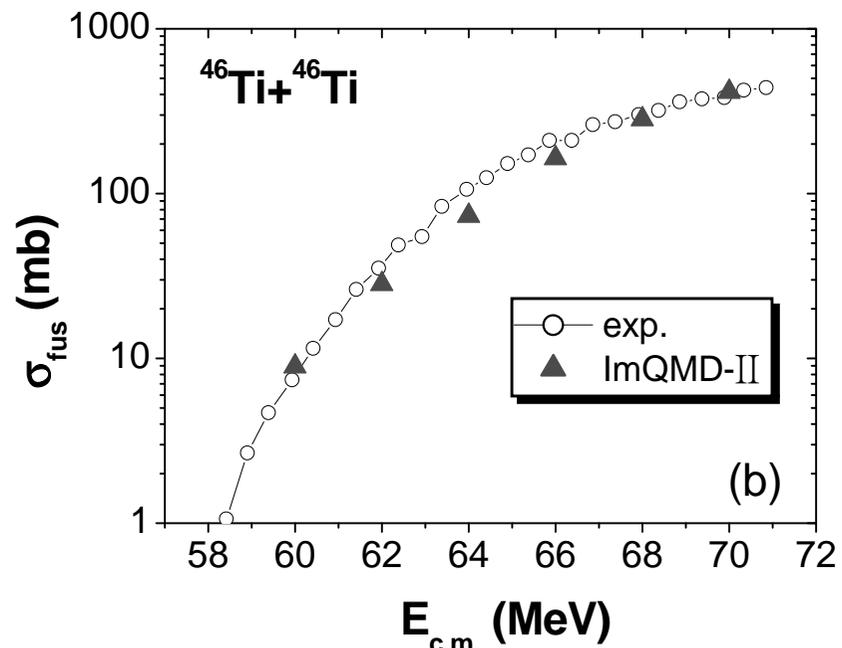
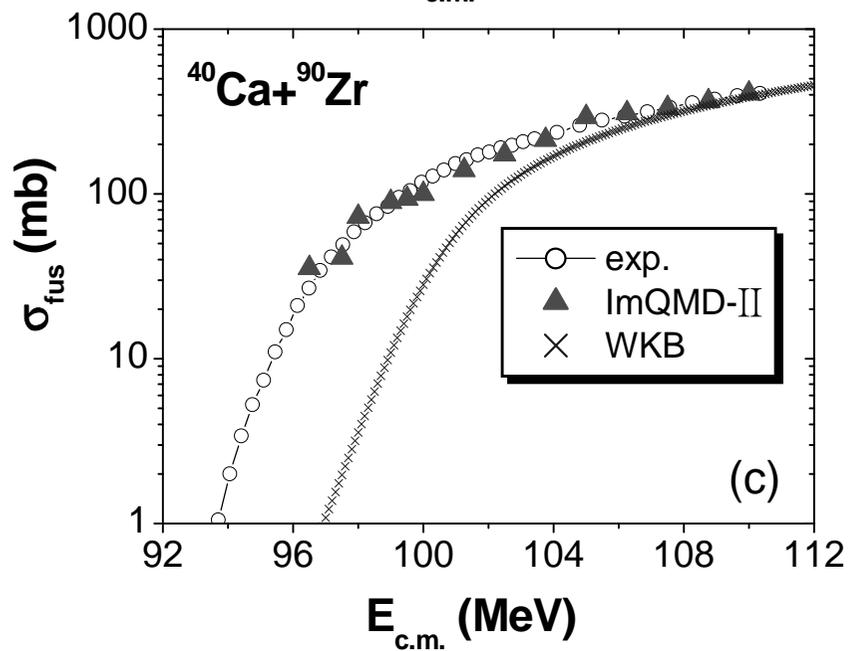
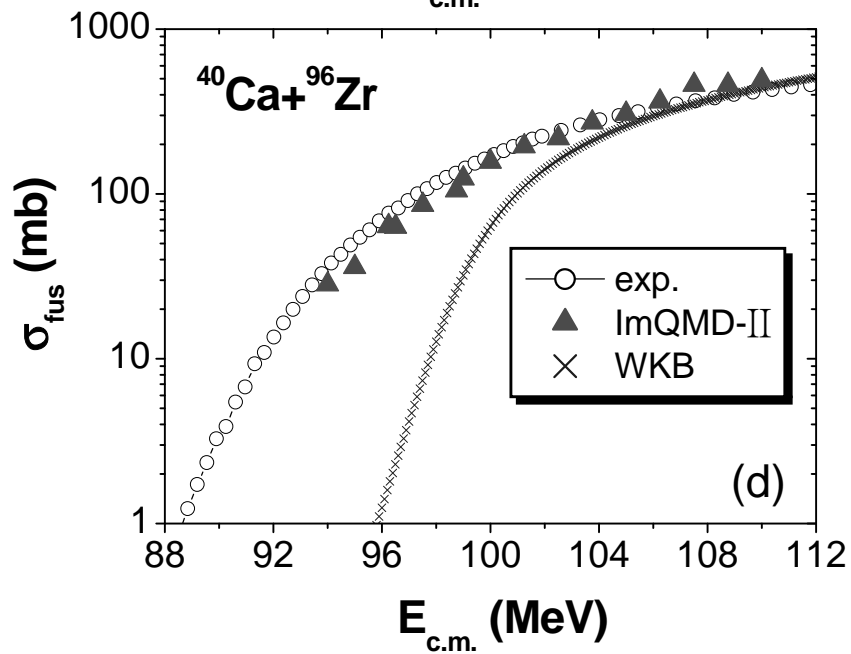
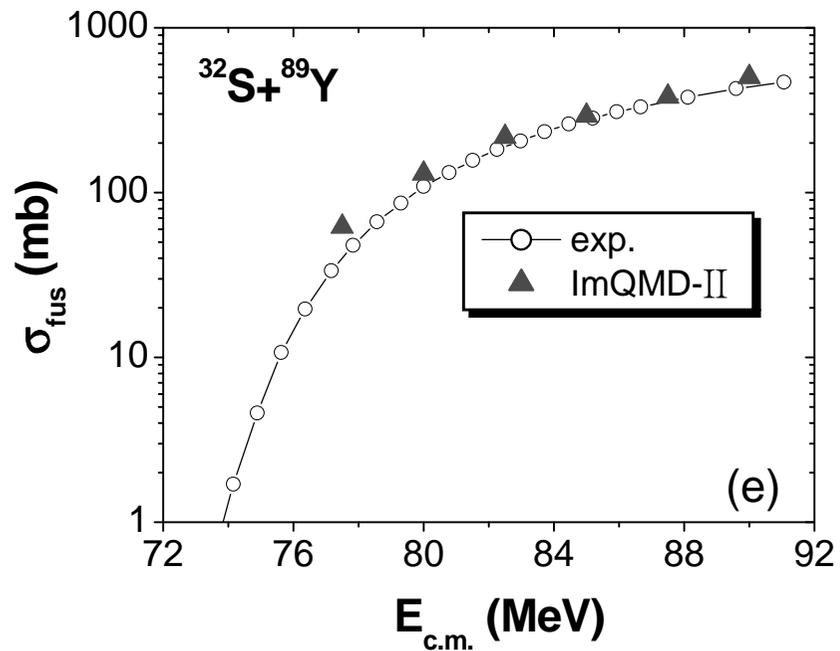
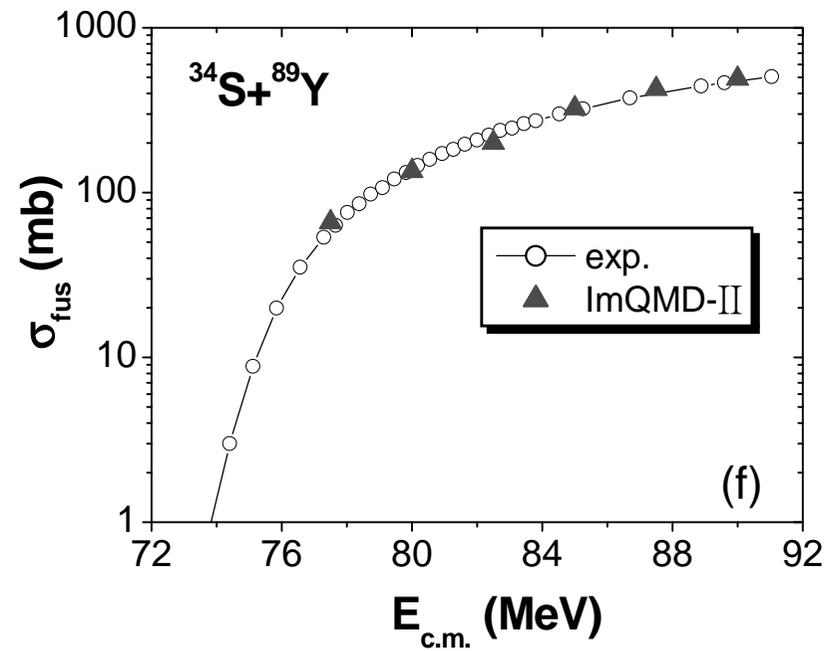
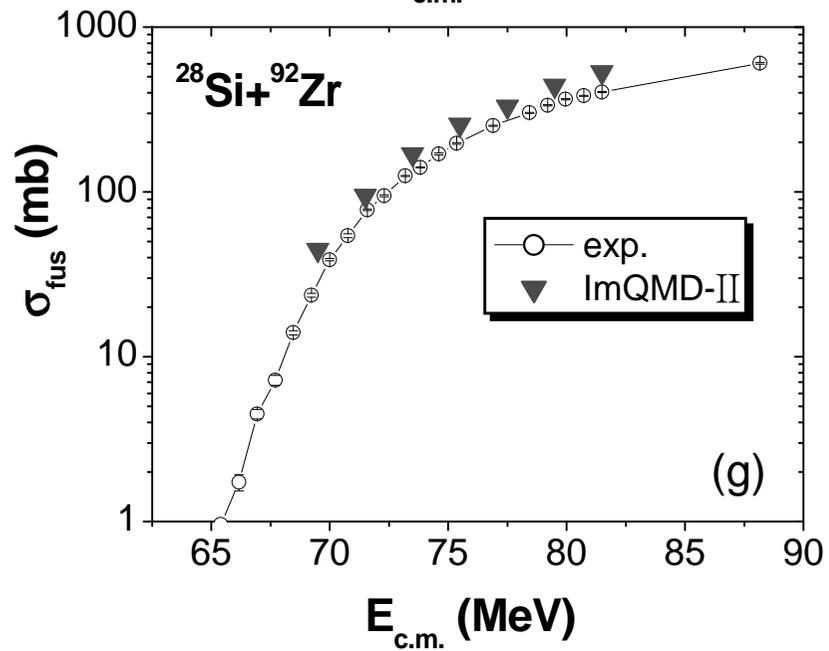
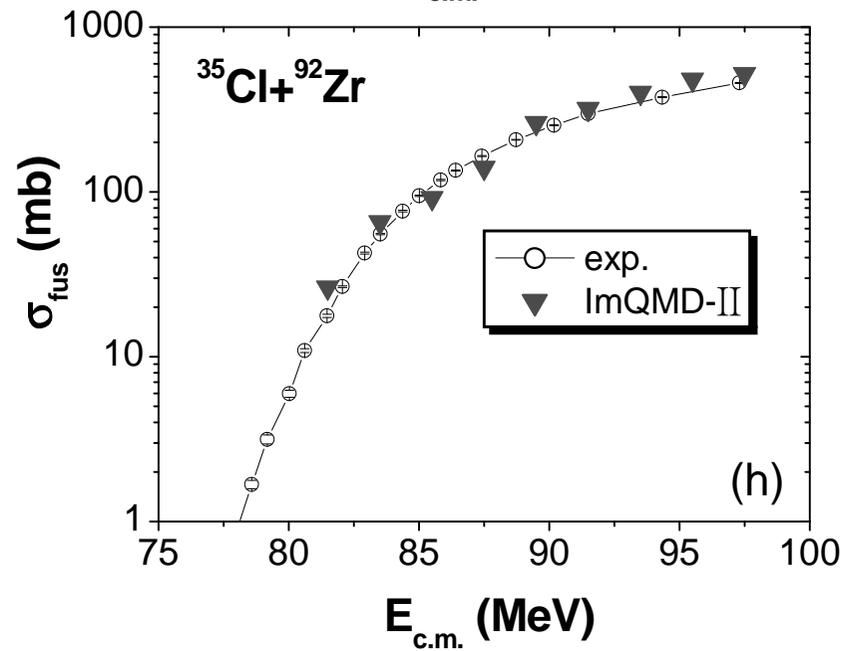
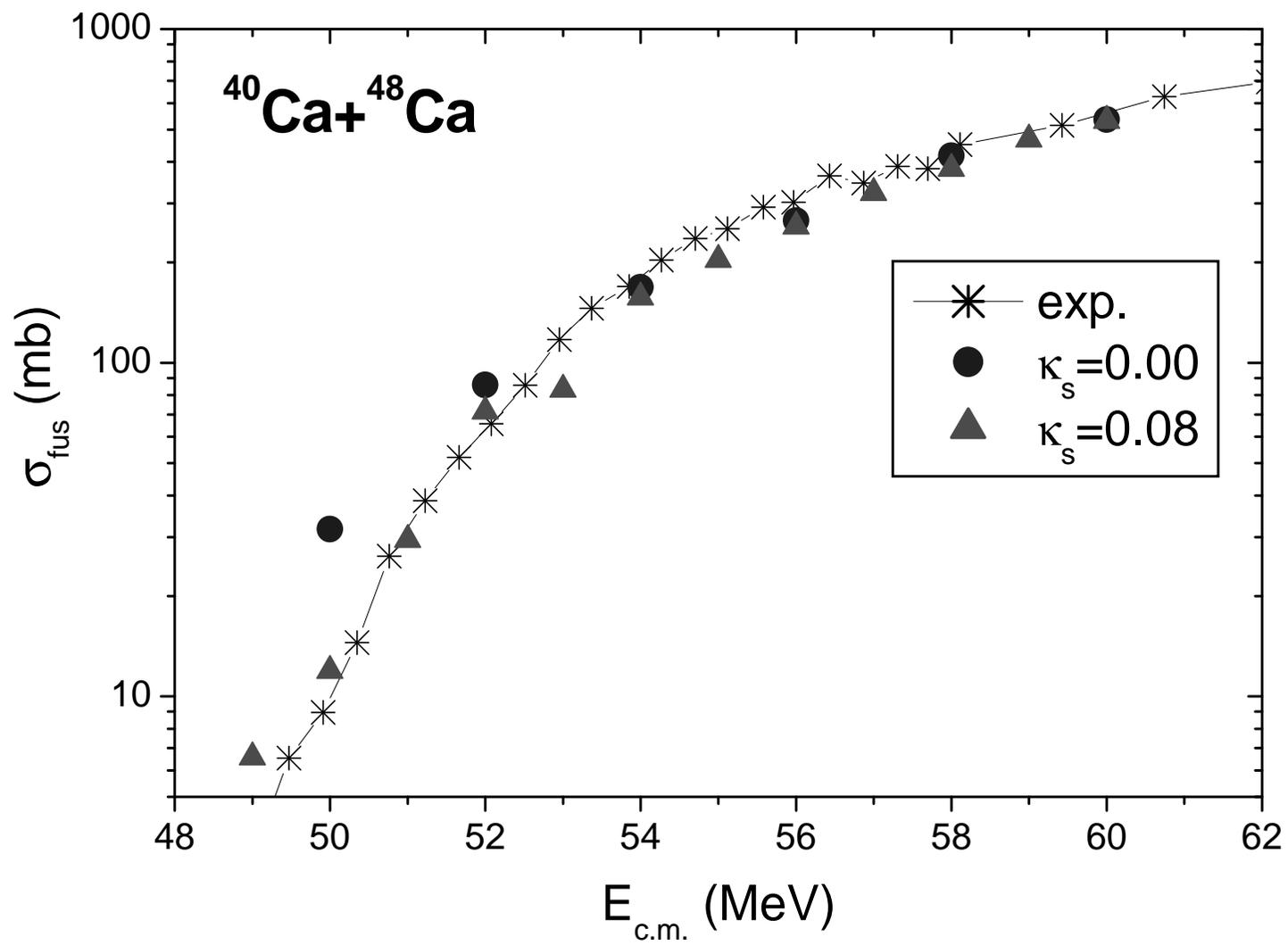
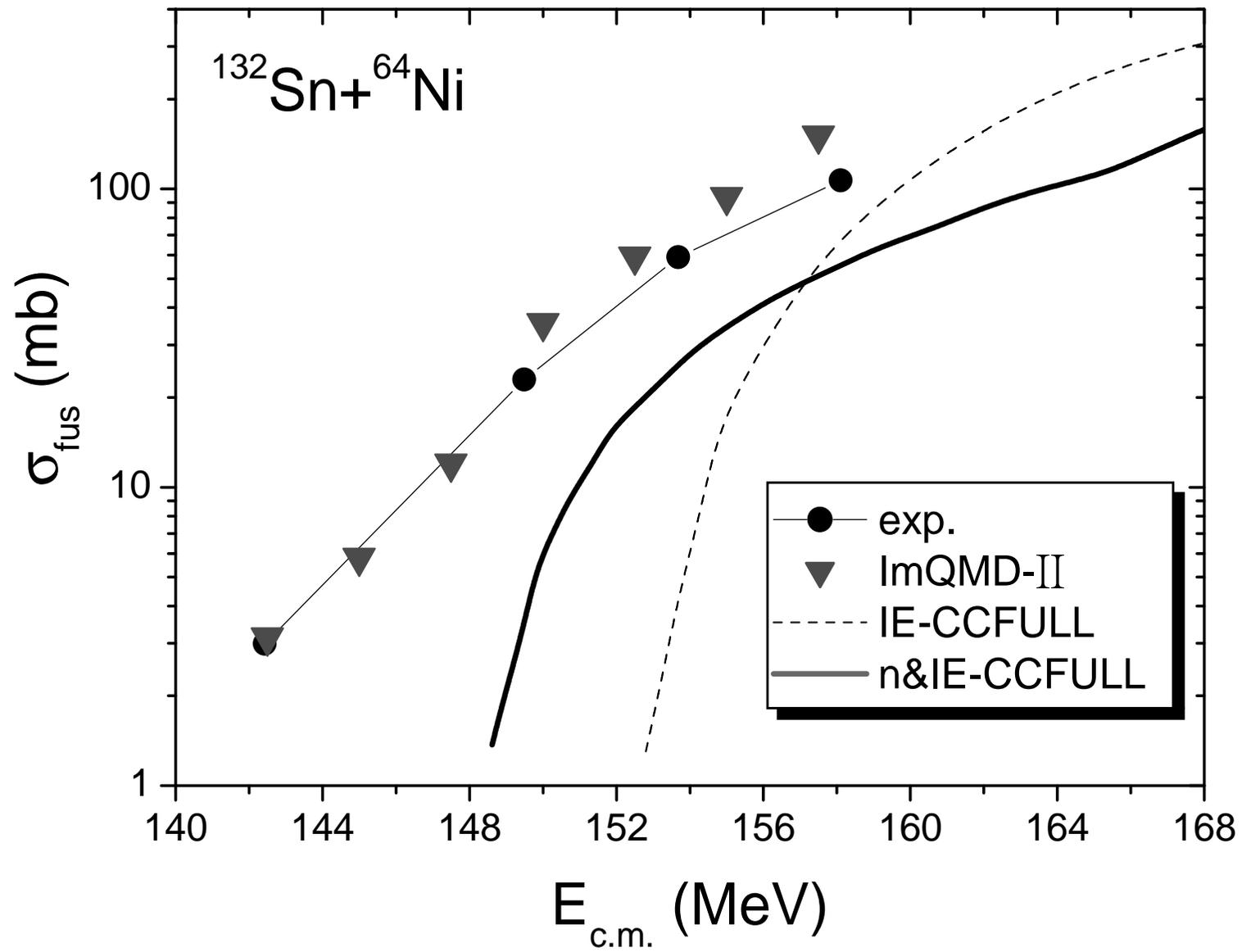
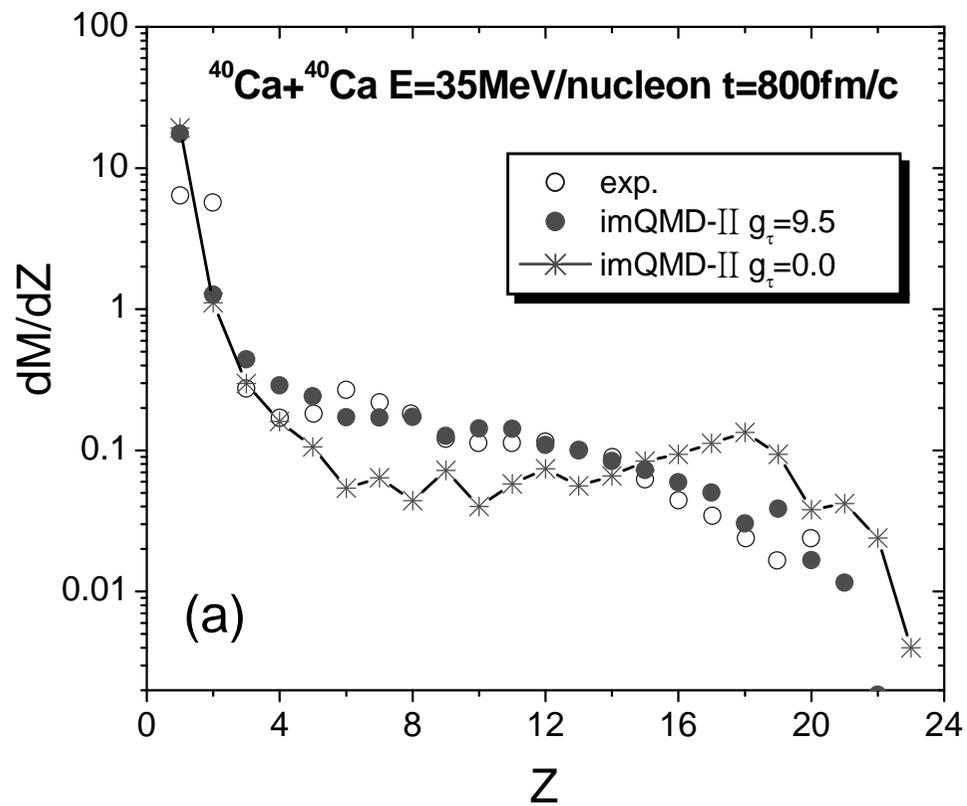
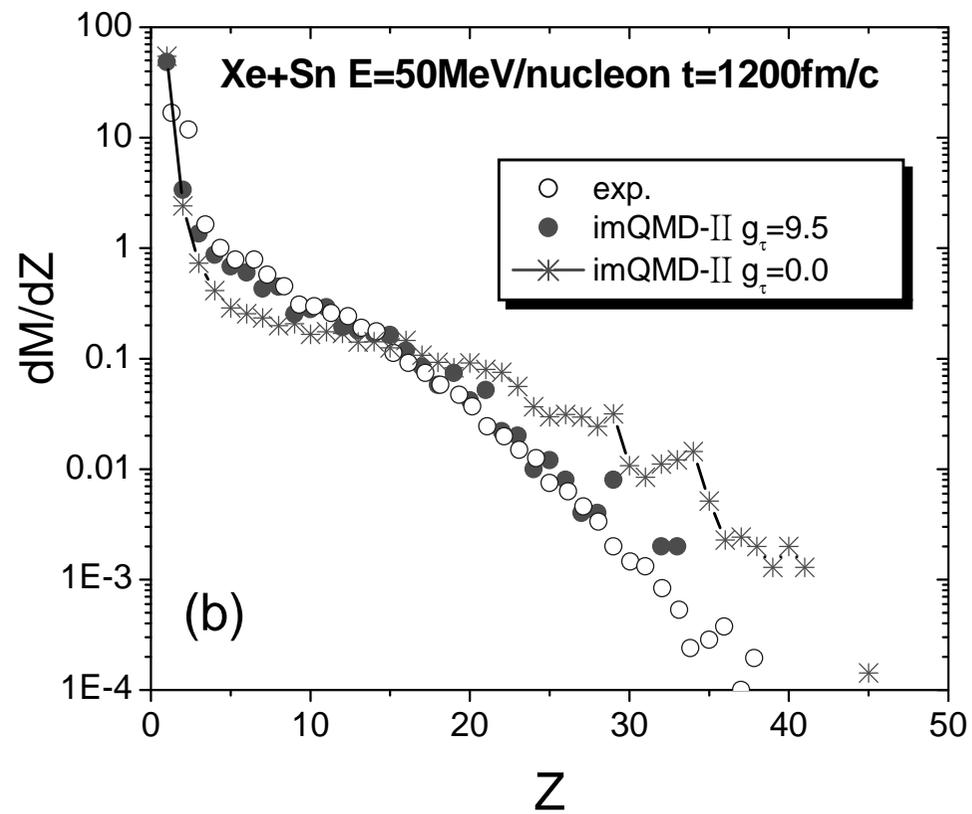
\end{document}